\documentclass[a4]{PoS}
\usepackage{amsbsy}
\usepackage{amsmath,amssymb,bbm}
\usepackage{epsfig}
\usepackage{euscript}

\def\rQCED{{\rm QCED}}

\newcommand{\FYFS}{F_{\mathrm{YFS}}}
\title{New developments in exact amplitude-based resummation in precision theory vs experiment}

\ShortTitle{Exact amplitude-based resummation}

\author{\speaker{B.F.L. Ward}\\
        Baylor University, Waco, TX, USA\\
        E-mail: \email{bfl\_ward@baylor.edu}}

\author{A. Mukhopadhyay\\
        IMSC, Chennai, IN\\
       E-mail: \email{aditi.banya@gmail.com}}

\abstract{We present recent developments in the application of exact amplitude-based resummation methods in the confrontation between precision theory and recent experimental results. As a consequence, we argue that these methods open the way to 1\% total theoretical precision in LHC and FCC physics when realized via MC event generators.}

\FullConference{12th International Symposium on Radiative Corrections (Radcor 2015) and LoopFest XIV (Radiative Corrections for the LHC and Future Colliders)\\
		15-19 June, 2015\\
		UCLA Department of Physics \& Astronomy Los Angeles, USA}

\centerline{BU-HEPP-16-01, Jan., 2016}

\begin{document}

\section{Introduction}
     In the 1988 ICHEP-Munich Conference Dinner, Prof. F. Berends and one of the authors (BFLW) considered the following question: "How accurate can exponentiation really be?" Would it limit or enhance the theoretical precision for a given level of exactness, LO, NLO, NNLO, ..., in the attendant fixed-order perturbative series? At that time, the context was the precision LEP physics program so that the focus was the SM EW theory. There were two main approaches on the market: the Jackson-Scharre (JS) approach~\cite{js} and the Yennie-Frautschi-Suura (YFS) approach~\cite{yfs}, where the latter was being pursued via MC event generator methods by Prof. S. Jadach, one of the authors (BFLW) and their collaborators~\cite{jadwrd}. One of us (BFLW) argued that with the JS approach there was a limit to the precision because one asserted that the overall exponential factor naively applied to all terms in the cross section when it is clear that this is not correct. In the YFS approach, an exact re-arrangement of the entire perturbative series is made, nothing is dropped, and there is no limit to how precise the result may be. As one can see in the SM Physics section of the 1989 Yellow Book~\cite{cern-89-08} edited by Prof. Berends, the discussion did bear some fruit. Today, an analogous discussion continues for the SM QCD theory.\par
     To wit, the ATLAS-CMS BEH boson discovery~\cite{beh,atlas-cms-2012} has ushered in the era of precision QCD, whereby we intend precision tags $\lesssim 1.0\%$ -- a new challenge for  both theory and experiment. Our response is exact amplitude-based resummation`\cite{irdglap1,irdglap2} realized on an event-by-event basis via parton shower/ME matched MC's. This realization gives enhanced precision for a given level of exactness in the attendant ME.Current realizations are in the Herwig6.5~\cite{hwg6} environment in the MC Herwiri1.031~\cite{herwiri}(IR-improved DGLAP-CS~\cite{dglap,cs} LO shower MC), in MC@NLO/Herwiri1.031(IR-improved NLO shower/ME matched MC) in the MC@NLO framework~\cite{mcatnlo}), in the new IR-improved DGLAP-CS Pythia8.183~\cite{pythia8} presented in Ref.~\cite{iri-pythia8}, and more recently in MG5\_aMC@\\
NLO/IRI-Pythia8.212((IR-improved NLO shower/ME matched MC)~\cite{elswh} in the MG5\_aMC@NLO\\
~\cite{mg5amc} framework. From comparisons with the ATLAS, CMS, D0 and CDF data~\cite{atlas,cms,D0,cdf} in Ref.~\cite{herwiri}, we see that the IR-improved Herwiri1.031 has better precision compared to the unimproved Herwig6.5 when $|\eta_a|,\;a=\ell,\;\bar{\ell},$ is in the central region in $Z/\gamma^*$ production and decay to lepton pairs $\ell, \bar{\ell}$. Here, we extend this analysis to the more forward LHCb data~\cite{lhcbdata} in the regime 
$2.0<|\eta_{\ell,\bar{\ell}}| <4.5$ and present a new paradigm for the next step in realizing our approach to precision QCD theory.\par
     The discussion is organized as follows. In the next section, we give a brief review of the parton shower implementation of exact amplitude-based resummation theory. Section 3 shows the comparisons with LHCb data and brings in the comparisons between other approaches and the LHC data. Section 4 then reviews Field's analysis~\cite{rick} of Drell-Yan at NLO to show the need to resum the IR limit. Section 5 shows how to tame the respective offending +-functions semi-analytically. Section 6 brings in the need to define what we mean by precision in the context of our discussion which is also closed therein.\par
\section{Review of Parton Shower Implementation of Exact Amplitude-Based Resummation Theory}
     The master formula in the theory, which applies both to the resummation of the reduced cross section
and to that of the evolution of the parton densities,
may be identified as
\begin{eqnarray}
&d\bar\sigma_{\rm res} = e^{\rm SUM_{IR}(QCED)}
   \sum_{{n,m}=0}^\infty\frac{1}{n!m!}\int\prod_{j_1=1}^n\frac{d^3k_{j_1}}{k_{j_1}} \cr
&\prod_{j_2=1}^m\frac{d^3{k'}_{j_2}}{{k'}_{j_2}}
\int\frac{d^4y}{(2\pi)^4}e^{iy\cdot(p_1+q_1-p_2-q_2-\sum k_{j_1}-\sum {k'}_{j_2})+
D_\rQCED} \cr
&\tilde{\bar\beta}_{n,m}(k_1,\ldots,k_n;k'_1,\ldots,k'_m)\frac{d^3p_2}{p_2^{\,0}}\frac{d^3q_2}{q_2^{\,0}},
%\end{split}
\label{subp15b}
\end{eqnarray}\noindent
where $d\bar\sigma_{\rm res}$ is either the attendant reduced cross section
$d\hat\sigma_{\rm res}$ or the attendant differential rate associated to a
DGLAP-CS~\cite{dglap,cs} kernel involved in the evolution of the corresponding parton densities $\{F_j\}$ and 
where the {\em new} (YFS-style~\cite{yfs,jadwrd}) {\em non-Abelian} residuals 
$\tilde{\bar\beta}_{n,m}(k_1,\ldots,k_n;k'_1,\ldots,k'_m)$ have $n$ hard gluons and $m$ hard photons and we show the final state with two hard final
partons with momenta $p_2,\; q_2$ specified for a generic $2f$ final state for
definiteness. The infrared functions ${\rm SUM_{IR}(QCED)},\; D_\rQCED\; $
are defined in Refs.~\cite{irdglap1,irdglap2,qced} as follows:
\begin{eqnarray}
{\rm SUM_{IR}(QCED)}=2\alpha_s\Re B^{nls}_{QCED}+2\alpha_s{\tilde B}^{nls}_{QCED}\cr
D_\rQCED=\int \frac{d^3k}{k^0}\left(e^{-iky}-\theta(K_{max}-k^0)\right){\tilde S}^{nls}_{QCED}
\label{irfns}
\end{eqnarray}
where the dummy parameter $K_{max}$ is such that nothing depends on it and where we have introduced
\begin{eqnarray}
B^{nls}_{QCED} \equiv B^{nls}_{QCD}+\frac{\alpha}{\alpha_s}B^{nls}_{QED},\cr
{\tilde B}^{nls}_{QCED}\equiv {\tilde B}^{nls}_{QCD}+\frac{\alpha}{\alpha_s}{\tilde B}^{nls}_{QED}, \cr
{\tilde S}^{nls}_{QCED}\equiv {\tilde S}^{nls}_{QCD}+{\tilde S}^{nls}_{QED}.
\end{eqnarray} 
Here, the superscript $nls$ denotes that the infrared functions are DGLAP-CS synthesized as explained in Refs.~\cite{dglpsyn,qced,irdglap1,irdglap2} and the infrared functions
$B_A,\; {\tilde B}_A,\; {\tilde S}_A, \; A=QCD,\; QED,$ may be found 
in Refs.~\cite{yfs,jadwrd,irdglap1,irdglap2,qced}. 
Note that the 
simultaneous resummation of QED and QCD large IR effects is exact here.\par 
    Via  their shower subtracted analogs, the new non-Abelian residuals $\tilde{\bar\beta}_{m,n}$ 
allow rigorous shower/ME matching. To achieve this
in (\ref{subp15b}) we make the replacements~\cite{qced}
\begin{equation}
\tilde{\bar\beta}_{n,m}\rightarrow \hat{\tilde{\bar\beta}}_{n,m}.
\end{equation}
The $\hat{\tilde{\bar\beta}}_{n,m}$ have had all effects in the showers
associated to the parton densities $\{F_j\}$ removed from them.\par
One may see how we make
contact between the $\hat{\tilde{\bar\beta}}_{n,m}$ and the
differential distributions in MC@NLO as
follows. We represent the MC@NLO differential cross section 
via~\cite{mcatnlo} 
\begin{equation}
\begin{split}
d\sigma_{MC@NLO}&=\left[B+V+\int(R_{MC}-C)d\Phi_R\right]d\Phi_B[\Delta_{MC}(0)+\int(R_{MC}/B)\Delta_{MC}(k_T)d\Phi_R]\nonumber\\
&\qquad\qquad +(R-R_{MC})\Delta_{MC}(k_T)d\Phi_Bd\Phi_R
\label{mcatnlo1}
\end{split}
\end{equation}
where $B$ is Born distribution, $V$ is the regularized virtual contribution,
$C$ is the corresponding counter-term required at exact NLO, $R$ is the respective
exact real emission distribution for exact NLO, $R_{MC}=R_{MC}(P_{AB})$ is the parton shower real emission distribution
so that the Sudakov form factor is 
$$\Delta_{MC}(p_T)=e^{[-\int d\Phi_R \frac{R_{MC}(\Phi_B,\Phi_R)}{B}\theta(k_T(\Phi_B,\Phi_R)-p_T)]},$$
where as usual it describes the respective no-emission probability.
The respective Born and real emission differential phase spaces are denoted by $d\Phi_A, \; A=B,\; R$, respectively.
We may note further that the representation of the differential distribution
for MC@NLO in (\ref{mcatnlo1}) is an explicit realization of the compensation 
between real and virtual divergent soft effects discussed in the 
Appendices of Refs.~\cite{irdglap1,irdglap2} in establishing the validity of 
(\ref{subp15b}) for QCD -- all of the terms on the RHS of (\ref{mcatnlo1}) are 
infrared finite. Indeed,
from comparison with (\ref{subp15b}) restricted to its QCD aspect we get the identifications, accurate to ${\cal O}(\alpha_s)$,
\begin{equation}
\begin{split}
\frac{1}{2}\hat{\tilde{\bar\beta}}_{0,0}&= \bar{B}+(\bar{B}/\Delta_{MC}(0))\int(R_{MC}/B)\Delta_{MC}(k_T)d\Phi_R\\
\frac{1}{2}\hat{\tilde{\bar\beta}}_{1,0}&= R-R_{MC}-B\tilde{S}_{QCD}
\label{eq-mcnlo}
\end{split}
\end{equation}
where we defined~\cite{mcatnlo} $$\bar{B}=B(1-2\alpha_s\Re{B_{QCD}})+V+\int(R_{MC}-C)d\Phi_R$$ and we understand here
that the DGLAP-CS kernels in $R_{MC}$ are to be taken as the IR-improved ones
which we have derived from (\ref{subp15b}) Refs.~\cite{irdglap1,irdglap2} and realized in Refs.~\cite{herwiri}, as we exhibit below. 
Here for simplicity of notation the QCD virtual and real infrared functions
$B_{QCD}$ and $\tilde{S}_{QCD}$ respectively are written without the superscript $nls$
and they are understood to be DGLAP-CS synthesized as explained in Refs.~\cite{irdglap1,irdglap2,qced} so that we
avoid doubling counting of effects. We observe that, in view of 
(\ref{eq-mcnlo}), 
the way to the extension of frameworks such as MC@NLO to exact higher
orders in $\{\alpha_s,\;\alpha\}$ is therefore open via our $\hat{\tilde{\bar\beta}}_{n,m}$. This way
will be taken up elsewhere~\cite{elswh}, as will be the extension of our methods to the POWHEG approach~\cite{powheg}.\par
     A key observation is that the relationship between the $\hat{\tilde{\bar\beta}}$ and the NLO corrections implies that the 
theoretical precision tag on the respective cross section requires study of the latter. Additionally, we are met with the NLO soft and collinear divergence
structure which is regulated by +-functions summarizing the cancellation of real and virtual effects. How does this impact the attendant theoretical
precision tag? To proceed, we look at the Drell-Yan process for the LHCb data to probe a different regime of the phase space compared 
to the ATLAS and CMS data analyzed in Refs.~\cite{herwiri}.\par
\par
\section{Interplay of IR-Improved DGLAP-CS Theory and NLO Shower/ME Precision: Comparison with LHCb Data}
     The LHCb data~\cite{lhcbdata} on $Z/\gamma^*$ production and decay to lepton pairs probes the regime wherein each lepton pair member satisfies
$2.0<\eta< 4.5$. This should be compared to $|\eta|<2.4$ for the ATLAS~\cite{atlas} data and to $|\eta|<2.1$ for the CMS~\cite{cms} $\mu^+\mu^-$ data
and $e^+e^-$ $Z/\gamma^*$ $p_T$ spectrum analyzed in Ref.~\cite{herwiri}. For the CMS~\cite{cms} $e^+e^-$ $Z/\gamma^*$ rapidity spectrum analyzed in Ref.~\cite{herwiri}, one lepton had $|\eta|<2.5$ and the other had $|\eta|<4.6$. We see that the LHCb data probe in all cases a new phase space regime compared to what we analyzed in our
studies in Refs.~\cite{herwiri}. Any complete treatment of theoretical precision has to address the situation in the entirety of the phase space measured at the LHC. \par
    We consider the LHCb results~\cite{lhcbdata} on the $Z/\gamma^*$ rapidity, $\phi_\eta^*$ and $p_T$ spectra in turn in Figs.~1-3.
\begin{figure}[h]
\begin{center}
%x\epsfig{file=pent-1.eps,width=140mm}
%\includegraphics[width=100mm]{fig-doe-2012-2a.eps}
\setlength{\unitlength}{0.1mm}
\begin{picture}(1600, 930)
\put( 400, 650){\makebox(0,0)[cb]{\bf (a)} }
\put(1240, 650){\makebox(0,0)[cb]{\bf (b)} }
\put(   -50, 0){\makebox(0,0)[lb]{\includegraphics[width=90mm]{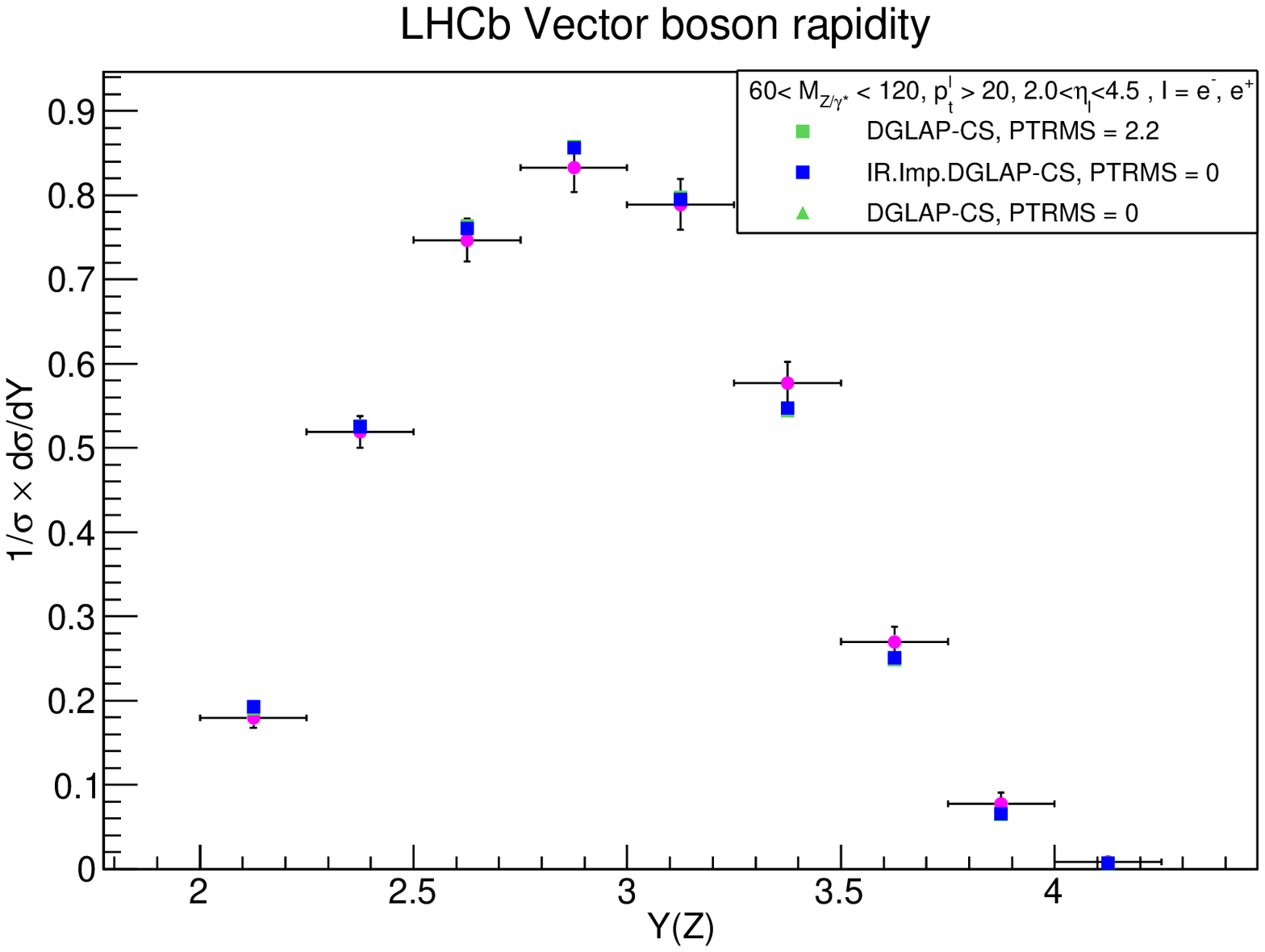}}}
\put( 800, 0){\makebox(0,0)[lb]{\includegraphics[width=90mm]{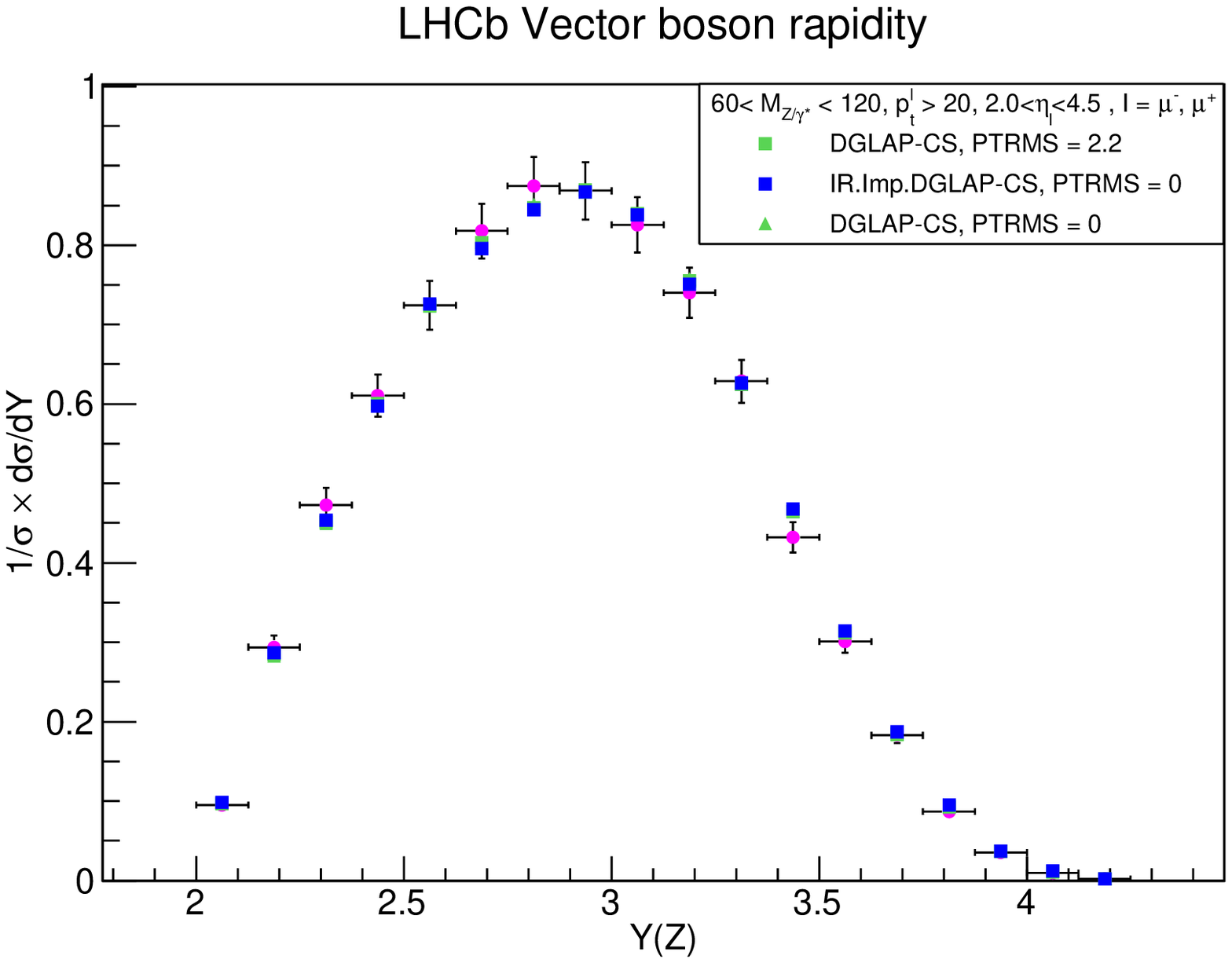}}}
\end{picture}
\end{center}
\caption{\baselineskip=8pt Comparison with LHCb data: (a), LHCb rapidity data on
($Z/\gamma^*$) production to $e^+e^-$ pairs, the circular dots are the data, the green(blue) squares are MC@NLO/HERWIG6.510($\rm{PTRMS}=2.2$ GeV/c)(MC@NLO/HERWIRI1.031); 
(b), LHCb rapidity data on ($Z/\gamma^*$) production to (bare) $\mu^+\mu^-$ pairs, with the same graphical notation as that in (a). In both (a) and (b), the green triangles are MC@NLO/HERWIG6.510($\rm{PTRMS}=$0). (In black and white, green = light, blue = dark.) These are otherwise untuned theoretical results. 
}
\label{figlhcb1}
\end{figure} 
\begin{figure}[h]
\begin{center}
%x\epsfig{file=pent-1.eps,width=140mm}
\includegraphics[width=100mm]{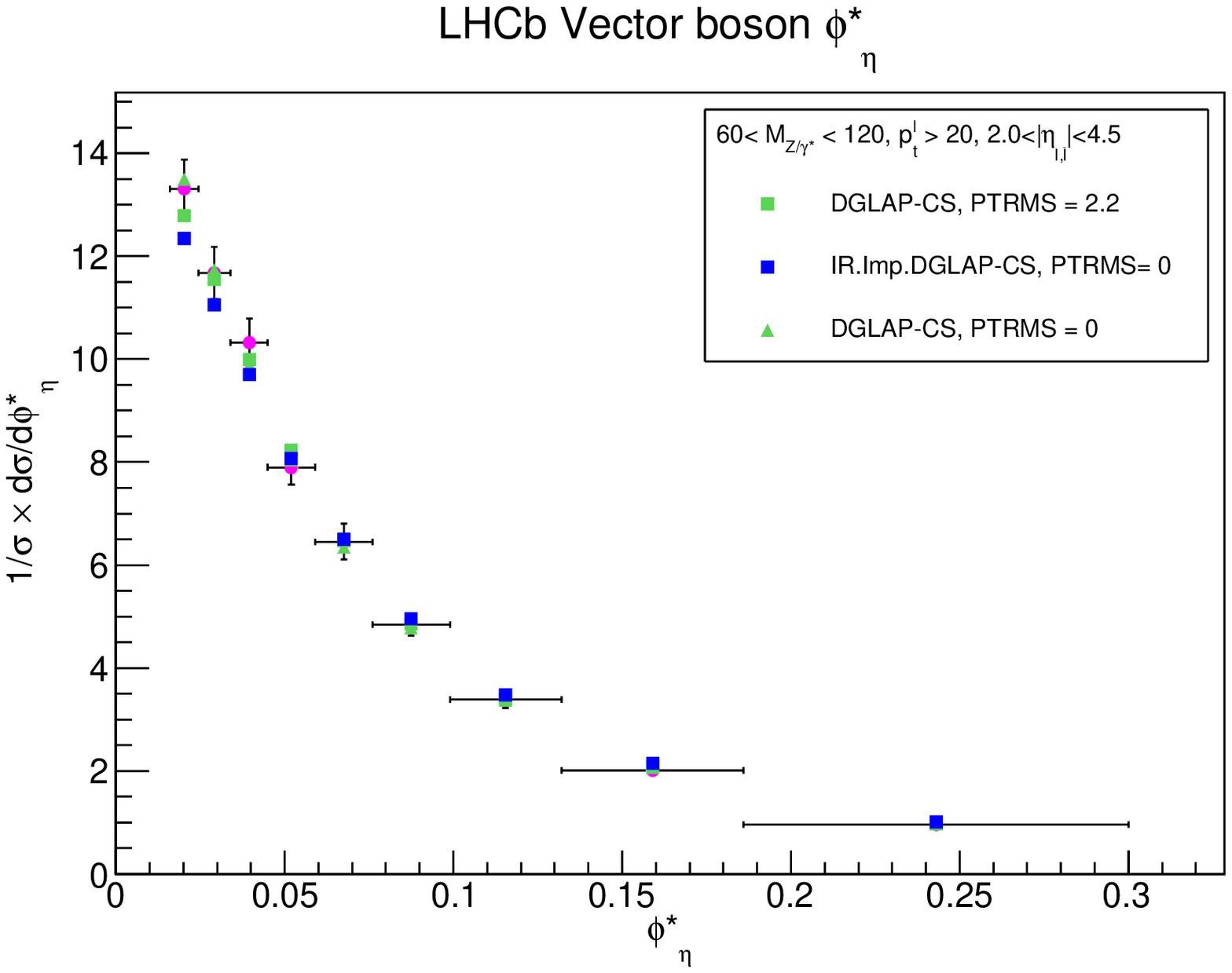}
\end{center}
\caption{\baselineskip=8pt Comparison with LHCb data on $\phi_\eta^*$ for the $\mu^+\mu^-$ channel in single $Z/\gamma^*$ production at the LHC. The legend (notation) for the plots is the same as in Fig.~1.} 
\label{figlhcb2}
\end{figure}
\begin{figure}[h]
\begin{center}
%x\epsfig{file=pent-1.eps,width=140mm}
%\includegraphics[width=100mm]{lhcb-pt-2011-mu-intm-3-15.eps}
\includegraphics[width=100mm]{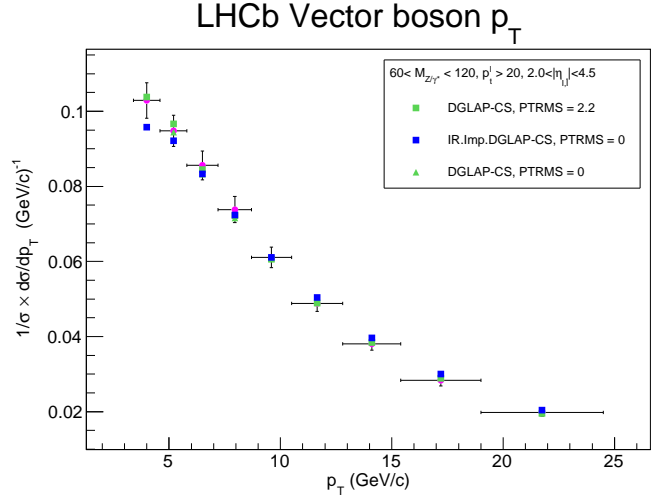}
\end{center}
\caption{\baselineskip=8pt Comparison with LHCb data on $p_T$ for the $\mu^+\mu^-$ channel in single $Z/\gamma^*$ production at the LHC. The legend (notation) for the plots is the same as in Fig.~1.} 
\label{figlhcb3}
\end{figure}     
What we see is that MC@NLO/Herwiri1.031 results in these figures are good fits to these more forward data without the need of an ad hocly hard intrinsic $p_T$
in the proton wave function. The two sets of MC@NLO/Herwig6.5 results, those with and those without a 2.2GeV/c intrinsic $p_T$ for the proton constituents, are also good fits to the LHCb data. These comparisons are in agreement
with our studies of the same rapidity variable in the more central regime in Refs.~\cite{herwiri}. The respective $\chi^2/d.o.f$ are 0.746, 0.814, 0.836 for the $e^-e^+$-Y data, 0.773, 0.555, 0.537 for the $\mu^-\mu^+$-Y data, 1.2, 0.23, 0.35 for the $\phi_\eta^*$ data, and 0.789, 0.183, 0.103 for the $p_T$ data, respectively for the MC@NLO/Herwiri1.031, MC@NLO/Herwig65(PTRMS=0) and MC@NLO/Herwig65(PTRMS = 2.2GeV/c) predictions. We see from the results in Figs.~1-3 that a proper interpretation of the data requires control of both the physical and technical precision of the theoretical predictions.\par
In the connection with precision, we note the discussions in Refs.~\cite{herwiri,1407-7290} of the similar comparisons between the FNAL and LHC data for other calculations in the literature~\cite{fewz,banfi,resbos}. These comparisons
show the need for a theoretical baseline analysis for precision studies. For example, the FEWZ~\cite{fewz} exact NNLO results undershoot the ATLAS data~\cite{atlas-moriond} on $\phi_\eta^*$ by $\sim 10\%$. To fromulate the respctive baseline, we turn next to Field's NLO analysis~\cite{rick} of Drell-Yan processes.\par

\section{Field's Analysis of Drell-Yan at NLO}
\begin{figure}[h]
\begin{center}
%x\epsfig{file=pent-1.eps,width=140mm}
\includegraphics[width=170mm]{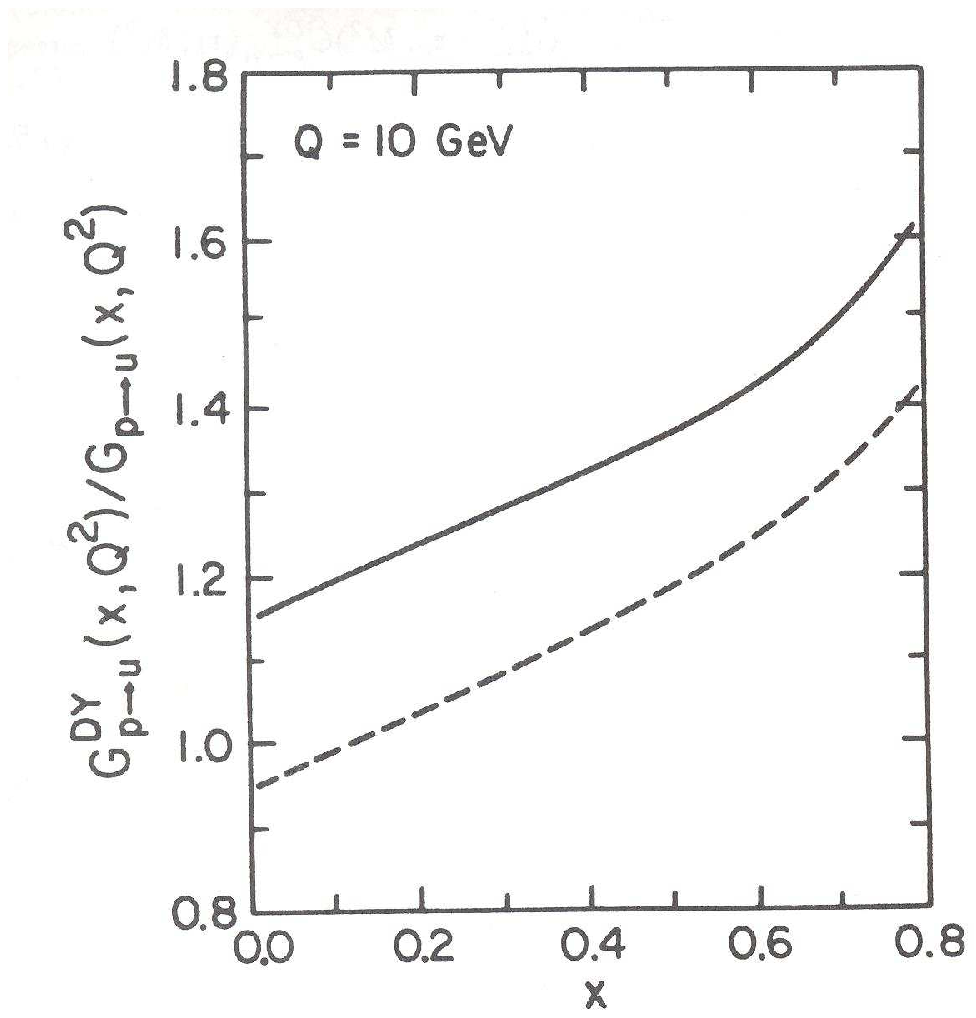}
\end{center}
\vspace{-140mm}
\caption{\baselineskip=11pt The ratio of the u-quark probability distribution
defined in the Drell-Yan process to that defined from the ${\rm F}_2$ structure function defined in deep inelastic lepton-nucleon scattering as discussed in Ref.~\cite{rick} at $Q= 10$ GeV, where the solid (dashed) curve corresponds to including (excluding) the total cross section in the attendant Drell-Yan distribution.}
\label{fig-rick}
\end{figure}

     To set up the baseline semi-analytical 
framework for theoretical precision 
estimates, we look at the NLO analysis of Drell-Yan processes in Ref.~\cite{rick} which we illustrate here in Fig.~\ref{fig-rick}. 
It is shown in eq.(5.5.30) in Ref.~\cite{rick} that
\begin{equation}
\frac{G^{DY}_{p\rightarrow q}(x,Q^2)}{G_{p\rightarrow q}(x,Q^2)}\operatornamewithlimits{\rightarrow}_{x\rightarrow 1}1+\frac{2\alpha_s(Q^2)}{3\pi}\ln^2(1-x)
\label{rickeq1}
\end{equation}
where $G^{DY}_{p\rightarrow q}$ ($G_{p\rightarrow q}$) is the respective Drell-Yan(DIS) structure function~\cite{rick} in a standard type of notation. 
No observable data, at LHC or the new FCC,
can have such behavior, which calls into question what a precision tag could even mean here?
\par

\section{Taming +-Functions in Drell-Yan at NLO and NNLO}
With the objective of taming what we see in Fig.~\ref{fig-rick} and (\ref{rickeq1}), we apply our master formula (\ref{subp15b}) to the NLO Drell-Yan formula of Refs.~\cite{guido-mart,humpvn}(We treat one flavor of unit charge for the $\gamma^*$ component only for reasons of pedagogy.) to obtain the IR-improved semi-analytical result
\begin{equation}
\begin{split}
\frac{d\sigma^{DY}_{res}}{dQ^2}&=\frac{4\pi\alpha^2}{9sQ^2}\int_0^1\frac{dx_1}{x_1}\int_0^1\frac{dx_2}{x_2}\big\{\left[q^{(1)}(x_1)\bar{q}^{(2)}(x_2)+(1\leftrightarrow 2)\right]2\gamma_qF_{YFS}(2\gamma_q)(1-z_{12})^{2\gamma_q-1}e^{\delta_q}\\
&\quad \times \theta(1-z_{12})\big[ 1+\gamma_q -7C_F\frac{\alpha_s}{2\pi}+ (1-z_{12})(-1+\frac{1-z_{12}}{2})\\
&\quad +2\gamma_q(-\frac{1-z_{12}}{2}-\frac{z_{12}^2}{4}\ln{z_{12}})\\
&\quad + \alpha_s(t)\frac{(1-z_{12})}{2\gamma_q}f^{DY}_q(z_{12})\big]\\
&\quad +\left[(q^{(1)}(x_1)+\bar{q}^{(1)}(x_1))G^{(2)}(x_2)+(1\leftrightarrow 2)\right] \\
&\quad \times \gamma_G F_{YFS}(\gamma_G)e^{\frac{\delta_G}{2}}[\alpha_s(t)\theta(1-z_{12})\big(\frac{t}{2\pi\gamma_G}(\frac{1}{2}(z_{12}^2(1-z_{12})^{\gamma_G}+(1-z_{12})^2z_{12}^{\gamma_G}))\\
&\quad +f^{DY'}_G(z_{12})/\gamma_G\big)]\big\}
\end{split}
\label{guido-eq5}
\end{equation}
where we have used the notation of Refs.~\cite{rick,guido-mart,humpvn}
for the hard corrections $\{f^{DY}_A, A=q,G\}$ and have introduced here
\begin{equation}
\begin{split}
\alpha_s f^{DY'}_G(z)&=\frac{\alpha_s}{2\pi}\frac{1}{2}[(z^2(1-z)^{\gamma_G}+(1-z)^2z^{\gamma_G})\ln\frac{(1-z)^2}{z}-\frac{3}{2}z^2(1-z)^{\gamma_G}+z(1-z)^{\gamma_G}\\
&\quad +\frac{3}{4}((1-z)^{\gamma_G}+z^{\gamma_G})],\\
\end{split}
\label{guido-eq6}
\end{equation}
and the following exponents and YFS infrared function, $\FYFS$, already needed for the IR-improvement of DGLAP-CS theory in Refs.~\cite{irdglap1,irdglap2}:
\begin{align}
\gamma_q &= C_F\frac{\alpha_s}{\pi}t=\frac{4C_F}{\beta_0}, \qquad \qquad
\delta_q =\frac{\gamma_q}{2}+\frac{\alpha_sC_F}{\pi}(\frac{\pi^2}{3}-\frac{1}{2}),\nonumber\\
\gamma_G &= C_G\frac{\alpha_s}{\pi}t=\frac{4C_G}{\beta_0}, \qquad \qquad
\delta_G =\frac{\gamma_G}{2}+\frac{\alpha_sC_G}{\pi}(\frac{\pi^2}{3}-\frac{1}{2}),\nonumber\\
\FYFS(\gamma)&=\frac{e^{-{C_E}\gamma}}{\Gamma(1+\gamma)}.
\label{resfn1}
\end{align}
We define $\beta_0=11-\frac{2}{3}n_f$ for $n_f$ active flavors 
in a standard way and $\Gamma(w)$ is Euler's gamma function of the complex variable $w$.
Note that we have mass factorized in (\ref{guido-eq5}) and (\ref{guido-eq6}) 
as it is done by Ref.~\cite{rick}.
It can be seen immediately that the regime at $z_{12}\rightarrow 1$ is now under control in (\ref{guido-eq5}):  we now get the
behavior such that the $\ln^2(1-x)$ on the RHS of (\ref{rickeq1}) is replaced by
$$\frac{2(1-x)^{\gamma_q}\ln(1-x)}{\gamma_q} - \frac{2(1-x)^{\gamma_q}}{\gamma_q^2},$$
and this vanishes for $x\rightarrow 1$.
This means that the hard correction now has the possibility to be rigorously compared 
{\em exclusively} to the data in a meaningful way. The extension 
of (\ref{guido-eq5}) to the NNLO results in Ref.~\cite{fewz,dynnlo} is also open. This is under study~\cite{elswh}.
We note that MC@NLO and POWHEG do not tame the $z\rightarrow 1$ divergence discussed here, the former swaps NLO emission for parton shower(PS) emission in the limit and the latter retains the NLO emission 
in the first parton shower emission,
and both PS and NLO emissions diverge in the limit. The two frameworks do tame the $p_T\rightarrow 0$ limit by the Sudakov effect.\par
   In Ref.~\cite{mg5amc}, in eqs.(2.126)-(2.129), it is noted that, if 
$\Delta=1+{\cal O}(\alpha_s)$ and $\Delta\rightarrow 0$ in IR limits,
precision is preserved by (see Ref.~\cite{mg5amc} for symbol definitions) 
$${ d\sigma^{({\mathbbm H})}_{ij}=\left(d\sigma^{(\text{NLO,E})}_{ij}-d\sigma^{(\text{MC})}_{ij}\right)\Delta,}\;\text{{ and}}\;
{ d\sigma^{({\mathbbm S})}_{ij}=d\sigma^{(\text{MC})}_{ij}\Delta+\sum_{\alpha=S,C,SC}d\sigma^{(\text{NLO},\alpha)}_{ij}+d\sigma^{(\text{NLO,E})}_{ij}(1-\Delta)}.$$ Our IR-improvement implies $ \Delta \propto (1-z)^{\gamma_A}, A=q,G$.
Such implementation is in progress~\cite{elswh}. 

\section{A Matter of Precision}
     A fundamental issue then obtains: What is the physical precision? In the
usual approach, one isolates the scales, renormalization, factorization, shower, ... , and varies them by $\frac{1}{\text{f}}$ to f, f~ $\sim 2,$
independently, correlatedly, ..., and a matter of taste enters.
The precision is taken from attendant variations of the observable. On another view~\cite{BLM}, scales should be determined by the dynamics of the process. Indeed,
from Refs.~\cite{zinn-shif}, typically, we have
\begin{align}Z&=\sum_k C_k(\frac{\alpha_s}{\pi})^k k^{b-1}A^{-k}k!\\
                    &=(-A\frac{\partial}{\partial A})^{b-1}\sum_k C_k(\frac{\alpha_s}{\pi A})^k k!,
\end{align}
with $C_k ={\cal O}(1)$ and $b,A$ process dependent.
From Ref.~\cite{whit-wat}, it follows that $Z$ is most probably asymptotic with
$Z-S_N = {\cal O}(\alpha_s/(\pi A))^N$, if $S_N$ is the sum of the first N terms. The error on $S_N \simeq \text{a factor $\lesssim$ 1 times}\;\; S_N-S_{N-1}$.
We have some experience from LEP~\cite{jad-prec}.\par
     We consider a known example:
\begin{align}
R&=\frac{\sigma(e^+e^-\rightarrow \text{hadrons})}{\sigma(e^+e^-\rightarrow \mu^+\mu^-)}\\
             &= R_{EW}[1+\sum_{n=1}^{\infty}c_n\left(\frac{\alpha_s(Q^2)}{\pi}\right)^{n} + \text{power corrections}],
\end{align}
where the { $c_n$} are known to { $n=4$} well enough to be in the PDG Review~\cite{pdg2014}.
%{\cpgr PDG2014: Chin. Phys. C{\bf 38}(2014)090001}. }
 We have the $\overline{MS}$ results~\cite{pdg2014}:
$c_1= 1,$~ $c_2=1.9857-.1152n_f,$~$c_3= -6.63694-1.20013n_f-.00518n_f^2-1.240\eta,$\\
for  $\eta=\frac{(\sum e_q)^2}{(3\sum e_q^2)},$ and
$c_4=-156.61+18.775n_f-.7974n_f^2 +.0215n_f^3 + (17.828-.575n_f)\eta, \;\cdots .$
Let us use these results to explore precision estimate methodology.\par
    As our ``toy case'', 
we set $Q=20$ GeV, $R_{EW}=1$, $n_f=5$.
We take QCD through { $n=(1),2$} as the predictions.
We take the { ($n=2$ term) $+$ $n=3$ term $+$ $n=4$ term} as the missing 
higher order correction, respectively.
Two methods of estimating the physical precision are applied:
 (A), varying the scale between $\frac{1}{2}Q$ and $2Q$;
 (B), using $f\times$ the $n=(1),2$ contribution, $f\sim \frac{1}{2}$, as the physical precision error.\par

We have also the following~\cite{pdg2014} under $Q\rightarrow \mu_R$:
 $\bar{c}_1(\mu_R^2/Q^2)= c_1,$ 
$\bar{c}_2(\mu_R^2/Q^2)=c_2 +\pi b_0c_1\ln(\mu_R^2/Q^2),$ 
$\bar{c}_3= c_3+(2b_0c_2\pi+b_1c_1\pi^2)\ln(\mu_R^2/Q^2)
+b_0^2c_1\pi^2\ln^2(\mu_R^2/Q^2),\;\cdots .$
We use these results accordingly.\par

     For NLO$\equiv n=1$, the QCD correction is $\delta_{QCD}=0.0476$. By method (A), the error is $$\Delta(\delta_{QCD})=\begin{cases} +.0074 \\
-.0056 .\end{cases}$$ By method (B) it is 
{ $$\Delta(\delta_{QCD})=\pm 0.024 .$$} The actual value is
{ $$\Delta(\delta_{QCD})(HO)=0.0014,$$ where HO denotes the higher order correction defined above.} Both methods give conservative estimates at NLO.\par

   At NNLO$\equiv n=2$, the QCD correction is $\delta_{QCD}=0.0508 .$ By method (A), the error is
$$\Delta(\delta_{QCD})=\begin{cases} +.00045 \\
-.0016 .\end{cases}$$ By method (B) it is 
$$\Delta(\delta_{QCD})=\pm 0.0016 .$$ The actual value is
$$\Delta(\delta_{QCD})(HO)=-0.0018 .$$ We see that one of the scale variations has nothing to do with the missing HO corrections. Method (B) gives an error
that is consistent with those HO corrections. Thus, the approach embodied
in Refs.~\cite{BLM,zinn-shif} can also be used for error estimation.\par
    Our methodology is then summarized as follows: we use approach (B) to estimate physical precision; we use semi-analytical baseline vs MC to estimate technical precision. This is a re-realization of our LEP/SLC paradigm~\cite{jad-prec}.
We look forward to its exploitation. One of us (BFLW) thanks Prof. W. Lerche for support and hospitality from the CERN TH Unit while part of this work was done.
\par
%\section{Summary}
%    Precision theory necessitates control of both the IR limit ($z\rightarrow 1$) and the collinear limit ($p_T\rightarrow 1$). We now have control over both limits. Some new physics may hang in the balance at both the LHC and the FCC.\par

\end{document}